\documentclass[aps,reprint,prl,amsmath,amssymb,superscriptaddress,showpacs]{revtex4-1}
\usepackage{color}

\usepackage{graphicx}
\usepackage{subfigure}
\usepackage{bbold}
\usepackage{verbatim}
\usepackage{float}
\usepackage{dsfont}
\usepackage[breaklinks]{hyperref}
\usepackage[utf8]{inputenc}
\usepackage[T1]{fontenc}

\begin{document}
\title{Exotic Ising dynamics in a Bose-Hubbard model}
\author{Luis Seabra}
\author{Frank Pollmann}
\affiliation{Max-Planck-Institut f\"{u}r Physik komplexer Systeme, 01187 Dresden, Germany}

\date{\today}
\begin{abstract}   
We explore the dynamical properties of a one-dimensional Bose-Hubbard model, where two  bosonic species interact via Feshbach resonance. We focus on the region in the phase diagram which is described by an effective, low-energy ferromagnetic Ising model in both transverse and longitudinal fields. In this regime, we numerically calculate the dynamical structure factor of the Bose-Hubbard model using the time-evolving block decimation method. In the ferromagnetic phase, we observe both the continuum of excitations and the bound states in the presence of a longitudinal field. Near the Ising critical point, we observe the celebrated $E_8$ mass spectrum in the excited states. We also point out possible measurements which could be used to detect these excitations in an optical lattice experiment.
\end{abstract}

\maketitle

%%%%%%%%%%%%%%%%%%%%%%%%%%%%%%%%%%%%%%%%%%%%%%%%%%%%%%

The search for emergent excitations arising from strong correlations has been very fruitful over the past few decades, from fractional quasiparticles in the fractional quantum Hall effect~\cite{laughlin83}, to effective magnetic monopoles in spin ice~\cite{castelnovo08}.
The one-dimensional (1D) transverse-field Ising model is another famous case showing collective excitations~\cite{mussardo-book}. 
It is a paradigmatic model for quantum phase transitions, hosting a critical point with central charge $c=1/2$~\cite{sachdev-book}.
Zamolodichkov showed that, by perturbing such a critical theory with a $\mathbb{Z}_2$ symmetry-breaking field, eight massive particles emerge in the  excitation spectrum~\cite{zamolodchikov89}.
These particles are the hallmark of an underlying  $E_8$ continuous symmetry, 
a very complex
 symmetry group discovered in mathematics, which attracts much interest in a  wide community,
 see e.g. Ref.~\cite{vogan07}.
The ratio between the masses of the two lightest particles predicted by $E_8$ symmetry has been  recently observed 
 by Coldea et al. in
 neutron-scattering studies of the quantum magnet~CoNb$_2$O$_6$~\cite{coldea10}.

The quest for complex many-body phenomena and unconventional excitations has greatly benefited from advances in ultracold atomic gases~\cite{lewenstein07,bloch08}. 
By confining the atomic cloud to an optical lattice, it is now possible to explore strongly-interacting lattice models with an unprecedented degree of control and tunability~\cite{jaksch98}.
Important milestones include  the observation of the superfluid to Mott insulator transition in bosonic systems~\cite{greiner02} and  the antiferromagnetic 1D Ising transition on a tilted optical lattice~\cite{simon11}.
Due to their long coherence times and the possibility of tuning the parameters of the system, cold atoms in optical lattices also allow the study of non-equilibrium dynamics~\cite{cheneau12,trotzky12} -- which is usually very difficult in a condensed-matter setting. A promising research area is the use of multi-component atomic mixtures
as a route to effective magnetic models, see e.g. Refs.~\cite{duan03,kuklov03}.
More specifically, recent theoretical work has proposed mixtures of atoms and molecules near a Feshbach resonance~\cite{chin10} in a Mott state as a route to the 1D Ising model~\cite{bhaseen09,hohenadler10,bhaseen11}. 
Here, the Feshbach resonant coupling between different species acts as a tunable handle on quantum fluctuations.

In this paper, we study the low-energy dynamical properties of a 1D Bose-Hubbard model describing two different bosonic species coupled by Feshbach resonance in the Mott insulating regime.
We  obtain the low-energy spectrum of this model via an appropriate dynamical structure factor.
The characteristic signatures of the broken-symmetry and disordered phases are clearly observed.
We reach an excellent agreement with the excitation spectrum of the Ising model.
By tuning the bosonic system close to a perturbed $c=1/2$ critical point, 
its excitation spectrum reveals the signatures of  $E_8$~symmetry.

%%%%%%%%%%%%%%%%%%%%%%%

We consider the following pairing  Bose-Hubbard  Hamiltonian, previously studied in~
\cite{radzihovsky04,dickerscheid05,sengupta05,rousseau08,rousseau09,bhaseen09,hohenadler10,bhaseen11,ejima11,bhaseen12}, 
\begin{align}
\mathcal{H} = & \sum_{i\alpha} \epsilon_\alpha n_{i\alpha}-\sum_{i\alpha}
t_\alpha
(b_{i\alpha}^\dagger b^{\phantom\dagger}_{i+1\alpha}+{\rm H.c.}) \nonumber 
\\
+ & \sum_{i\alpha\alpha^\prime} \frac{U_{\alpha\alpha^\prime}}{2}n_{i\alpha}(n_{i\alpha^\prime}-\delta_{\alpha\alpha'}) +
g\sum_i(b_{im}^\dagger b^{\phantom\dagger}_{ia} b^{\phantom\dagger}_{ia}+{\rm H.c.}),
\label{eq:model-BH}
\end{align}
describing two species of bosons $b_{i\alpha}$ on a 1D lattice, where $n_{i\alpha}=b^\dagger_{i\alpha} b^{\phantom \dagger}_{i\alpha}$. Atoms are  labeled by $\alpha=a$ while molecules are labeled by $\alpha=m$.
Here $\epsilon_\alpha$ are on-site potentials, $t_\alpha$ are hopping parameters between nearest-neighbour sites, and  $U_{\alpha\alpha'}$ are on-site  interactions.
Two atoms  form a molecule via \mbox{$s$-wave} pairing, driven by Feshbach coupling~$g$.
The Feshbach interaction breaks the independent conservation of the number of atoms and molecules, but the total number \mbox{$N_T\equiv\sum_i(n_{ia}+2n_{im})$} is conserved. We work in the canonical ensemble, by keeping the total density $\rho_T=N_T/L$ fixed.

%%%%%%%%%%%%%%%%%%%%%%%%%%%%%%%%%%%%%

\begin{figure}[b!]
\centering
\includegraphics[width=8.3cm]{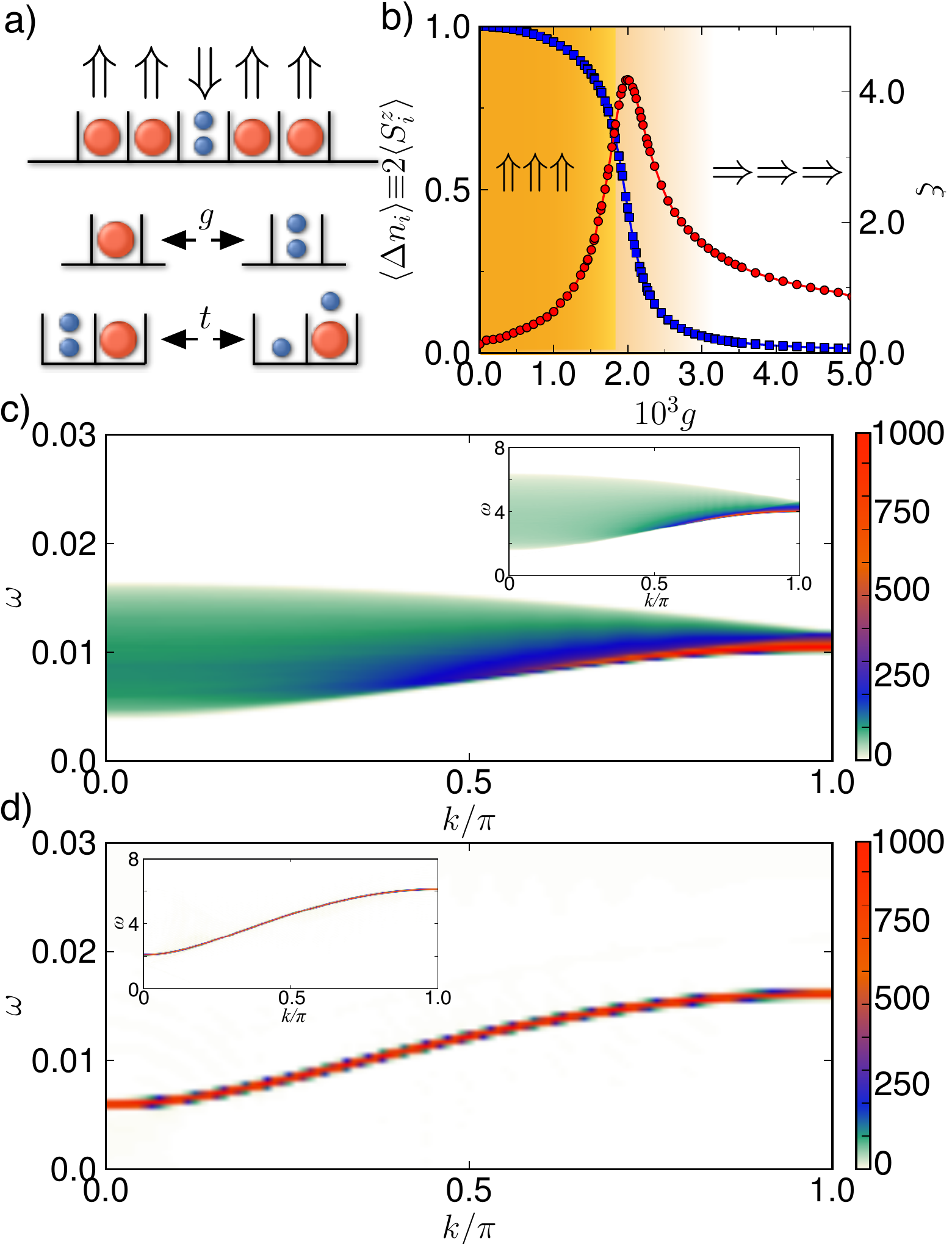} 
\caption{
(Color online)
(a) Mapping between the Bose-Hubbard model Eq.~(\ref{eq:model-BH}) in the second Mott lobe and the Ising chain Eq.~(\ref{eq:model-Ising}).  
Two atoms (one molecule)  map(s) to effective spin-down (spin-up).
A local excitation converts one molecule into two atoms, 
which  propagate via Feshbach coupling (transverse field $\Gamma$) and  boson hopping (Ising interaction $J$).
(b) Phase diagram of the full model Eq.~(\ref{eq:model-BH}) for $t_a$$=$$5.04\times10^{-2}$~\mbox{($J$$=$$0.01$)} and constraints detailed in the text, measured by the relative boson density. 
(c) Dynamical structure factor of the full model for $g$$=$$1.06\times10^{-3}$ ($\Gamma$$=$$0.3J$) and  $\Delta\epsilon_m$$=$$1.5 \times10^{-4}$ and (d) for \mbox{$g$$=$$3.54\times10^{-3}$} (\mbox{$\Gamma=J$}). Dashed lines are the dispersion minima of the Ising model for respective values of $\Gamma$, while
insets show $\mathcal{S}(k,\omega)$ of the Ising model, with the characteristic continuum of excitations~(c) and  quasi-particle dispersion~(d).}
\label{fig:figure-1}
\end{figure}

The low-energy behaviour of the Hamiltonian Eq.~(\ref{eq:model-BH}) in the Mott regime with $\rho_T=2$ can be conveniently described with the aid of an effective 1D quantum Ising model~\cite{bhaseen09,hohenadler10,bhaseen11},  which is also helpful in guiding us to the regions of interest.
The ``effective spin'' degrees of freedom are \mbox{$|\Uparrow\rangle\equiv|1;0\rangle$} and \mbox{$|\Downarrow\rangle\equiv|0;2\rangle$} in the occupation basis $|n_a;n_m\rangle$, see Fig.~\ref{fig:figure-1}(a). 
We truncate the Hilbert space to a maximum of  three atoms and one molecule per site, which is already a good approximation to canonical soft-core bosons for the large $U/t$ limit considered here~\cite{bhaseen11}.
This choice allows the hopping of atoms, even if a pair  is already  present on a site.
The effective Ising model (up to an additive constant) is  obtained 
via a strong-coupling expansion around the small-hopping limit
\begin{eqnarray}
\mathcal{H} \simeq -J\sum_{i} S^z_i S^z_{i+1} + h \sum_iS_i^z + \Gamma \sum_iS_i^x  + \mathcal{O}(t^3).
\label{eq:model-Ising}
\end{eqnarray}
The effective spin operators have a direct interpretation in terms of bosons,
\begin{eqnarray}
S^z_i= (n_{im}-n_{ia}/2 ) /2\equiv\Delta n_i/2
\end{eqnarray} 
 measures the  imbalance in the density of bosons at site~$i$, 
 while
 \begin{eqnarray}
S^x_i=1/(2\sqrt{2})[b_{im}^\dagger b_{ia}^{\phantom \dagger} b_{ia}^{\phantom \dagger} + b_{ia}^\dagger b_{ia}^\dagger b_{im}^{\phantom \dagger}]
\end{eqnarray}
 accounts for inter-species fluctuations.
At a qualitative level, the Ising exchange interaction $J$ arises from the motion of bosons, the longitudinal field $h$ tunes an overall imbalance between the two species, and the transverse field $\Gamma$ controls the fluctuations between the two species, cf. Fig.~\ref{fig:figure-1}(a).

%%%%%%%%%%%%%%%%%%%%%%%%%%%%%%%%%%%%%%%%%%%%
We find the ground state of the full Hamiltonian Eq.~(1) using a variant of the infinite density-matrix renormalization group (iDMRG) method \cite{mcculloch08, kjall13}, yielding a matrix-product-state (MPS) representation of the ground-state wave-function in the thermodynamic limit.
We find that matrix bond dimensions $\chi\lesssim 30$ are enough to describe the ground states studied, with a truncation error up to $10^{-10}$.
The  time-evolving block decimation (TEBD) method \cite{vidal03,vidal07} is then used to calculate a dynamical structure factor function $\mathcal{S}(k,\omega)$ of the bosonic model, following the general strategy laid out in Refs.~\cite{kjall11,milsted12,zauner12,phien12}.
%
%
%$\mathcal{S}(k,\omega)$
This function measures the response to fluctuations between the two species, corresponding to the $S^y$ operator in the Ising language:
\begin{eqnarray}
S^y_i=1/(2\sqrt{2i})[b_{im}^\dagger b_{ia}^{\phantom \dagger} b_{ia}^{\phantom \dagger} - b_{ia}^\dagger b_{ia}^\dagger b_{im}^{\phantom \dagger}].
\end{eqnarray}
The two-point dynamical correlation function
\begin{align}
 C(i,t)=\langle\psi_0|S^y_i(t)S^y_0(0)|\psi_0\rangle,
 \label{Cxt}
\end{align}
is calculated with a real-time evolution of the ground-state MPS $|\psi_0\rangle$ after $S^y$ is applied to a given site, and its Fourier transform yields $\mathcal{S}(k,\omega)$.
The sampled time is extended by extrapolating $C(i,t)$ with linear prediction~\cite{white08}.
 We stop the simulation once the ``light-cone'' of local correlations gets close to the boundary of a fixed window size (typically $L\approx200$), ensuring that our simulations do not suffer from finite-size effects~\cite{phien12}.
Since the low-energy dynamics are set by the effective model,  the ``light-cone'' and the entanglement entropy grow very slowly with time. 
Hence, we are able to reach extremely large times,  up to $t_{\sf max}\approx 10^4U_{aa}$,  while keeping the truncation error  down to 
$\lesssim1\times10^{-7}$ by setting $\chi_{\sf max} = 60$.
We have checked the convergence of our results with the Trotter time step $\Delta t$, settling on $\Delta t=0.1-0.2$.
%

%%%%%%%%%%%%%%%%%%%%%%%%%%%%%%%%%%%%%%%%%%%%%%%%%

The ground-state properties of the pairing Bose-Hubbard model have been studied recently~\cite{rousseau08,rousseau09,bhaseen09,hohenadler10,bhaseen11}. 
While, previously, emphasis was placed  
on the region of the phase diagram where the effective model is the antiferromagnetic Ising chain, here we focus on parameters yielding an effective  \emph{ferromagnetic} model.
In order to make contact with previous work, we impose the following constraints~: $t_a=2t_m$, $\epsilon_a=0$, $U_{am}=2U_{aa}$ and \mbox{$U_{aa}=2$} --- thus the energy scale is set by the on-site interaction. 
The parameters of the corresponding Ising Hamiltonian Eq.~(\ref{eq:model-Ising}) are then  found through the strong-coupling expansion~\cite{bhaseen11}  
\begin{align}
J & = \frac{63t_a^2}{16}, \label{eq:balance-1}\\
h & = -2+6t_a^2 + \epsilon_m + \Delta\epsilon_m, \label{eq:balance-2}\\
\Gamma & =  2g\sqrt{2}. \label{eq:balance-3}
\end{align}
We choose a representative value of $t_a=5.04\times10^{-2}$ \mbox{($J=0.01$)},  which avoids a slow down in the dynamics at lower values of $t_a$. 
Eq.~(\ref{eq:balance-2}) predicts that the effective longitudinal field $h$ is canceled by tuning  $\epsilon_m=1.9848$. 
However, even after this, we observe in our simulations the effects of the presence of an effective longitudinal field, which
arises from contributions in perturbation theory in small $t_a$ beyond the second-order expansion considered in Ref.~\cite{bhaseen11}. 
Rather than extending this calculation to higher order, we add an extra perturbation, $\Delta\epsilon_m$, to the molecular potential~in~Eq.~(\ref{eq:balance-2}) to compensate  for this ``stray'' longitudinal field. In general, this term depends on the values of $t_a$ and $g$ chosen.
Probably, this is also a more realistic way of achieving the same result in an experimental setting, since it just involves further fine-tuning of~$\epsilon_m$, instead of balancing a version of Eq.~(\ref{eq:balance-2}) with more, higher-order, terms.

%%%%%%%%%%%%%%%%%%%%%%%%%

%
The resulting phase diagram, where the Feshbach coupling $g$ is the only free parameter and the order parameter is $\Delta n\equiv$$ 2 S^z$,  is shown in Fig.~\ref{fig:figure-1}(b) for $\Delta\epsilon_m=0$.
For low values of $g$ an ordered phase based on the product state with one molecule localised per site, $|$$\Uparrow\rangle$, is found.
The stray longitudinal field  described above  biases the system towards this state, rather than the  ``atomic'' ordered state based on two atoms localised per site,  $|$$\Downarrow\rangle$.
All of the physical behaviour  described here still holds in the case where the on-site potentials have been tuned to favour the atomic ground state.
By increasing the value of $g$ the system goes through a crossover into a disordered phase.
This crossover is revealed by the non-diverging peak of the correlation length near $g=2\times10^{-3}$ ($\Gamma\approx J/2$), indicating that the Ising critical point is nearby.

The low-energy excitation spectrum is revealed by the transverse dynamical structure factor function $\mathcal{S}(k,\omega)$. 
The energy scales observed here are set by the effective Ising parameters and are hence rather low with $\omega\approx 0.01$.
This is well below the Mott gap $U_{aa}$, above which single-particle excitations appear.
We start with the ``molecular'' ground state where the $S^y$ excitation in Eq.~(\ref{Cxt}) dissociates one molecule into two atoms at site~$i$, i.e. flips $|\Uparrow\rangle\rightarrow|\Downarrow\rangle$.
Evolution in time creates a domain of atoms, which is a bound state of two domain walls propagating in opposite directions.
In the absence of a longitudinal field, the domain walls are deconfined as the two ground states are perfectly degenerate.
This is achieved by setting $g=1.06\times10^{-3}$ ($\Gamma=0.3J$) and fine tuning the on-site  potential  $\Delta\epsilon_m=1.5\times10^{-4}$.
The dynamical structure factor function, shown in Fig.~\ref{fig:figure-1}(c), displays a broad continuum of excitations around  $k=0$, sharpening  closer to the Brillouin zone edge at $k=\pi$.
The agreement with the pure Ising model is very good, as shown by the matching of the onset of the continuum in the Bose-Hubbard model with the energy of the lowest excitation (dashed line), of the transverse-field Ising chain with $\Gamma=0.3J$. The  Ising $\mathcal{S}(k,\omega)$ is also shown in the inset, see Ref.~\cite{mussardo-book,kjall11}.
A finite longitudinal field confines the domain walls, destroying the continuum of excitations, except for resonances at specific momenta and energies, which can be seen as massive ``meson'' bound states~\cite{mccoy78}.
An effective longitudinal field has the same effect here,  breaking up the continuum observed in Fig.~\ref{fig:figure-1}(c)  (not shown).

The dynamical structure factor in the disordered phase for large $g$    is shown in Fig.~\ref{fig:figure-1}(d) for a representative value \mbox{$g=3.54\times10^{-3}$} (\mbox{$\Gamma=J$}).
A well-defined quasi-particle with a quadratic dispersion is visible, which can be identified with the single spin-flip excitation of the Ising chain in the disordered phase, see dashed line  inset.
%

%%%%%%%%%%%%%%%%%%%%%%%%%%%%%%%%%%%%%

The most impressive feature of the Ising model is the $E_8$ symmetry~\cite{mussardo-book}, 
which is revealed when it is tuned to the critical point $\Gamma=J/2$ with a longitudinal field $|h|\ll|\Gamma|$ applied.
This is described by the perturbation of a $c=1/2$ conformal field theory, which is still integrable and leads to  exactly eight massive bound states, whose mass ratios are known analytically~\cite{zamolodchikov89}.
\begin{figure}[b!]
\centering
\includegraphics[width=8.8cm]{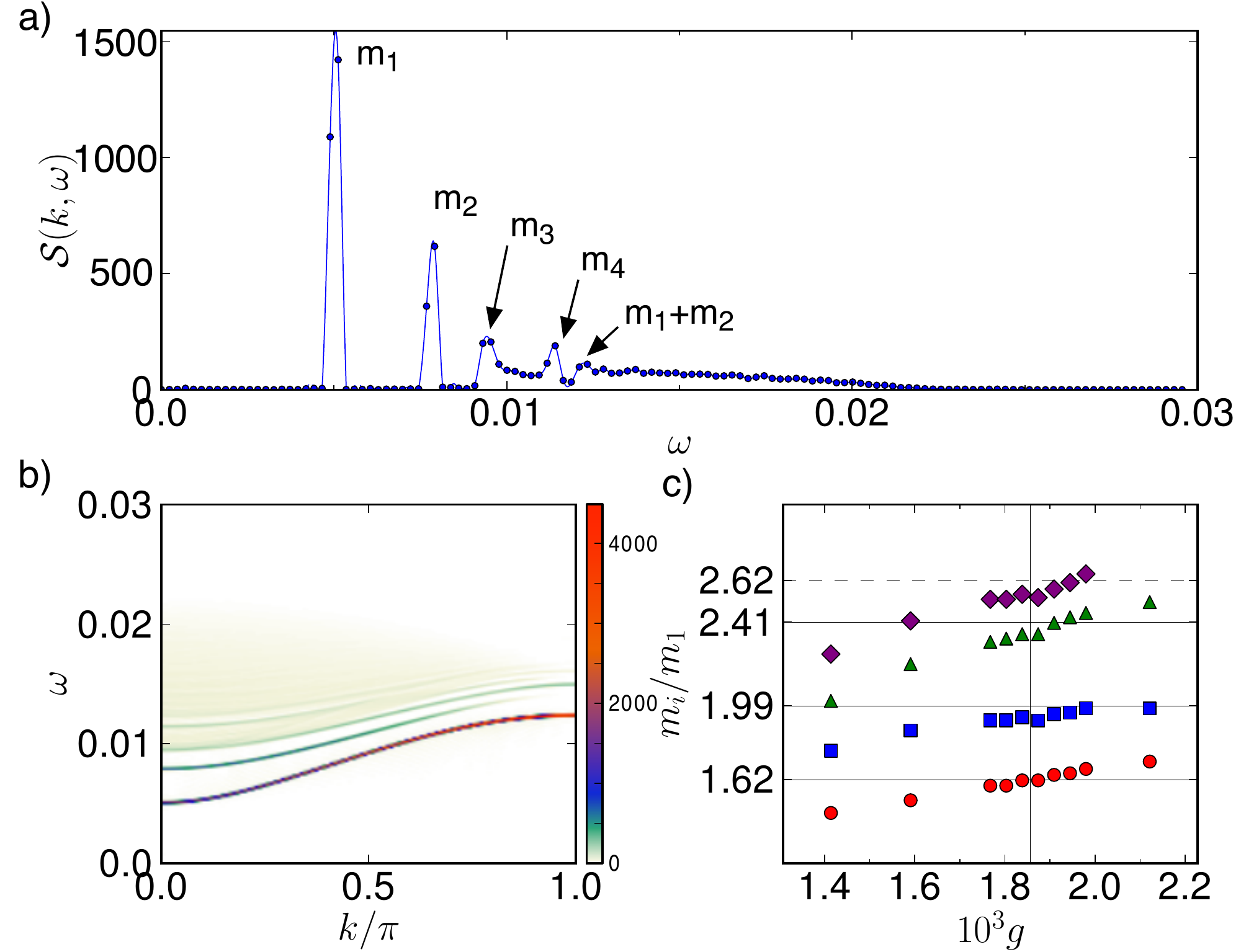}
\caption{
(Color online) Signatures of $E_8$ symmetry in a Bose-Hubbard model. (a) Cut at $k=0$ of the dynamical structure factor for $g=1.80\times10^{-3}$ ($\Gamma\approx J/2$) and \mbox{$\Delta\epsilon_m= -5\times10^{-4}$}.
 (b) Dynamical structure factor for the same parameters as (a), resolved in full momentum space.
(c) Relative masses of excitations to that of the lightest one, as a function of Feshbach coupling $g$, for fixed $\Delta\epsilon_m=-1\times10^{-3}$. Horizontal lines show the values analytically predicted from the Lie algebra~$E_8$ from Ref.~\cite{zamolodchikov89}, displayed in (a).
}
\label{fig:figure-2}
\end{figure}
 The mass ratio between  the two lightest particles is the golden ratio $(1 + \sqrt5)/2=1.618(...)$ and has been experimentally observed in CoNb$_2$O$_6$~\cite{coldea10}.
However, heavier excitations are difficult to measure there, since they are located within the continuum. Furthermore, it is not possible to modify this by tuning the longitudinal field, which is fixed~by~the~inter-chain coupling.
We now tune the parameters in our bosonic model to this interesting region.
The excitation spectrum of the system  close to the perturbed Ising critical point, with
$g=1.80\times10^{-3}$ (\mbox{$\Gamma=0.51J$}) and \mbox{$\Delta\epsilon_m= -5\times10^{-4}$} \mbox{$(|h|\approx|\Gamma|/10$)}, is revealed by examining  the dynamical structure factor function in Figs.~\ref{fig:figure-2}(a,b). 
At least five different excitations are clearly identified above the continuum, which we can associate with the first four particles of  the $E_8$ theory and the bound-state pair $m_1+m_2$. 
In  Fig.~\ref{fig:figure-2}(c) we present a sweep in Feshbach coupling near this point, while keeping the molecular potential perturbation fixed at $\Delta\epsilon_m=-1\times10^{-3}$ for better convergence of the results.
The mass ratios calculated in this region are very close to the analytically predicted ones (horizontal lines), crossing them for a value of $g\approx1.83\times10^{-3}$  ($\Gamma\approx0.52J$).
The observation of this highly non-trivial sequence in the energy spectrum is a clear evidence for the emergence of $E_8$ symmetry in a Bose-Hubbard model.
This small renormalisation of the critical value of $\Gamma$ probably arises from contributions in higher order from perturbation theory.
The different mass ratios increase roughly linearly with~$g$. The heaviest particles become progressively more difficult to observe with increasing $g$, since their spectral weight decreases.
The ratios obtained do not change significantly when  the molecular potential perturbation is doubled to  $\Delta\epsilon_m=-2\times10^{-3}$, staying within error bars of the data points in Fig.~\ref{fig:figure-2}(c).
 The stability of the results for such a large ratio of fields, $|h|/|\Gamma|\approx1/2$, matches that observed in the Ising chain~\cite{kjall11}. 
%

%%%%%%%%%%%%%%%%%%%%%%%%%%%%%%%%%%%%%%%%
%
Recent developments in Bragg spectroscopy applied to cold atoms
allow the study of the full excitation spectrum, resolved in momentum and energy, even in the presence of an optical lattice~\cite{clement09,ernst10}. 
A complementary, and simpler, scheme for observing some of the behaviour described here 
 is also desirable.
Motivated by recent experimental~\cite{sherson10} and theoretical work~\cite{kleine08a,honer12}, we now look at the propagation of a single excitation in space and time.
We prepare the system in the molecular ground state for  \mbox{$g=1.06\times10^{-3}$},
apply the $S^y$ dissociation excitation at a site $i_0$, and
 track the evolution in time of the relative local density $\Delta n$ at a site $i$.
\begin{figure}[h!]
\centering
\includegraphics[width=8.8cm]{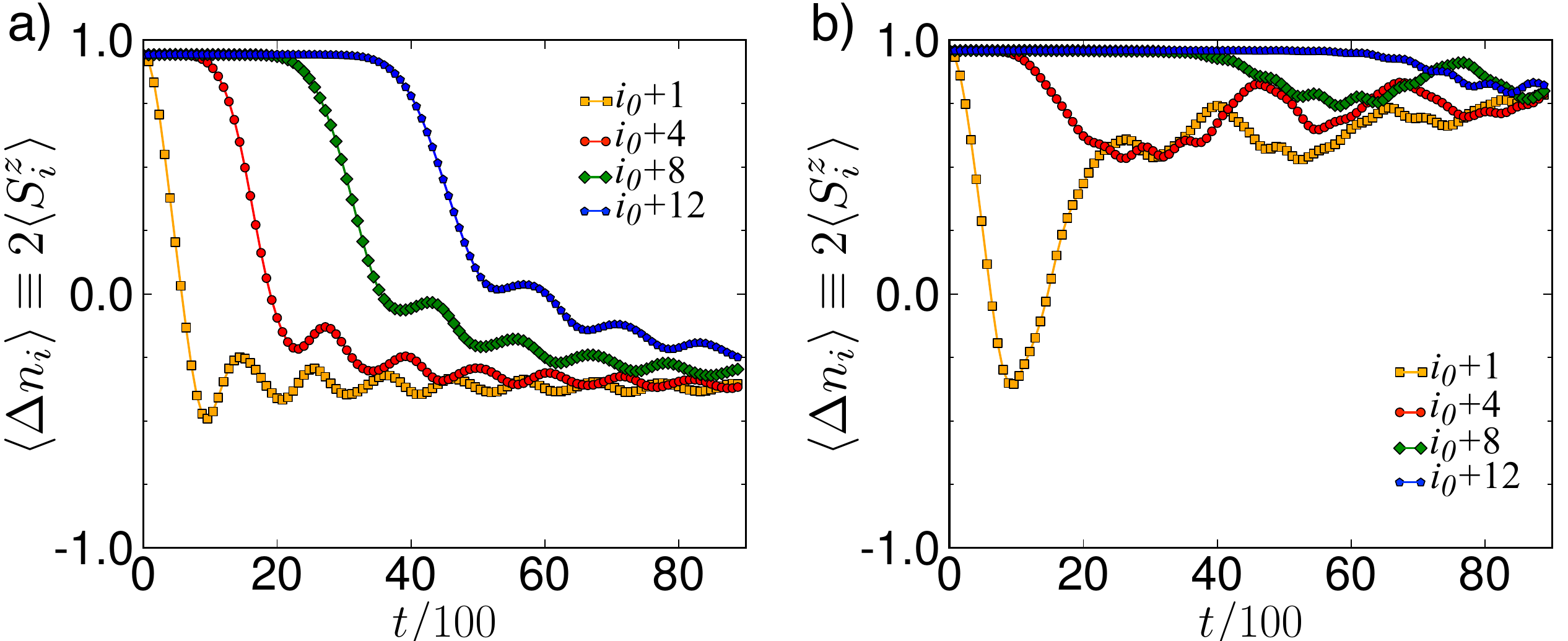}
\caption{
(Color online)
Time evolution of the relative boson density after a local quench dissociates two molecules into one atom on the molecular state for $g=1.06\times10^{-3}$.
(a) By tuning $\Delta\epsilon_m=1.5\times10^{-4}$, the effective longitudinal field is suppressed and the domain of atoms grows unbounded with time.
(b) By tuning $\Delta\epsilon_m=-1\times10^{-3}$, a longitudinal field is induced. The growth of the domain is thwarted and it quickly collapses, resulting in the confinement of the excitation.
}
\label{fig:figure-3}
\end{figure} 
The effective longitudinal field has been tuned to $h=0$ by setting \mbox{$\Delta\epsilon_m=1.5 \times 10^{-4}$} in Fig.~\ref{fig:figure-3}(a).
At a qualitative level,  $\Delta n_i$  drops suddenly  when a domain wall reaches site~$i$, signalling the dissociation of the molecule at that site ($\Delta n>0$)  into two atoms ($\Delta n<0$).
Since there is no effective longitudinal field to confine the  domain walls, the $\Delta n$$<$$0$ domain persists in the long-time limit, and it grows with more sites being progressively flipped, leading to the continuum of excitations in the dynamical structure factor shown in Fig.~\ref{fig:figure-1}(c). 
The behaviour in the presence of an effective longitudinal field, $\Delta\epsilon_m=-1\times10^{-3}$
, is shown in Fig.~\ref{fig:figure-3}(b).
The domain of atoms still grows up to a few sites away from $i_0$.  
However, the induced confining potential inhibits the propagation of the domain walls and the domain quickly
collapses.
The sites which were excited eventually return to a state with $\Delta n \approx 1$ in the long-time limit, which is different from the original molecular ground state.
%

%%%%%%%%%%%%%%%%%%%%%%%%%%%%%%%%%%%%

Systems of hundreds of cold atoms confined to 1D optical lattices have been extensively explored in the last decade~\mbox{\cite{bloch08,chin10}}, including
 heteronuclear bosonic mixtures~\cite{thalhammer08,catani09} which can be described by Bose-Hubbard models similar to Eq.~(\ref{eq:model-BH}), see e.g.~Ref.~\cite{dalmonte13}.
The key requirements to observe the behaviour  described here
are the formation of a symmetry-breaking insulating phase and the presence of fluctuations able to destroy it.  
The presence of a trap potential should not affect much the results, as long as the central insulating domain with $\rho_T=2$ is large enough.
In order to ensure an insulating phase, the ratios $t_\alpha/U_{\alpha\alpha}$, which depend on the lattice depth and the intra-species scattering length, should be~$\lesssim0.1$.
Biasing the on-site potentials $\epsilon_\alpha$, controls the effective longitudinal field $h$, and therefore the stability of the particles and their spectral weight relative to the continuum.
The Feshbach coupling $g$  is itself an effective term depending, among others, on the background scattering length and the width of the resonance~\cite{dickerscheid05,chin10}.
The effective Ising interaction  $J$ in Eq.~(\ref{eq:model-Ising}) is controlled by the ratios
between the different $U$, including $U_{am}$, which is the term freely adjustable in  Feshbach resonance experiments by the detuning of the magnetic field away from the resonance value~\cite{chin10,dalmonte13}.
In this way the $\Gamma/J$ and $h/J$ ratios could be controlled, allowing a sweep near the critical point as in Fig.~\ref{fig:figure-3}(c), in order to find the optimal set of parameters corresponding to $E_8$ symmetry.
%

%%%%%%%%%%%%%%%%%%%%%%%%%%%%%%%%%%%%

In conclusion, we have shown that the low-energy excitations of a 1D Bose-Hubbard pairing model can faithfully simulate the dynamical properties of the quantum Ising chain.
We find the characteristic Ising features in the low-energy spectrum, such as the incoherent continuum, quadratic quasi-particles, and, above all, massive excitations emerging from $E_8$ symmetry.
Recent developments in manipulating systems described by models such as Eq.~(\ref{eq:model-BH}), and in  accessing the low-energy excitation spectrum, open up the fascinating possibility of realising
this behaviour in a cold atoms~experiment.

%%%%%%%%%%%%%%%%%%%%%%%%%%%%%%%%%%%%%%%%%%%

We are grateful to M. J. Bhaseen and F. H. L. Essler for introducing us to this interesting model. We thank J. Eisert, S. Ejima, H. Fehske,  J. A. Kj\"{a}ll, A.~M. Turner  and M. Zaletel for fruitful discussions. 

%%%%%%%%%%%%%%%%%%%%%%%%%%%%%%%%%%%%%%%%%%%%%%%%%%%%%%

\end{document}